\documentclass[a4paper,12pt]{article}
\usepackage[utf8]{inputenc}
\usepackage{amsmath,subeqnarray,subfigure}
\usepackage{float,graphicx,textcomp,fontenc}
\usepackage{graphicx}
\usepackage{hyperref}

\def\H{\mbox{H}}

\def\sech{\,\mbox{sech}}

\title{Supersymmetric features of the Error and Dawson's functions}
\author{Marco A. Reyes and Rafael Arcos-Olalla
\\
{\small\it Departamento de F\'isica, Universidad de Guanajuato} \\
{\small\it Apdo. Postal E143, C.P. 37150 Le\'on, Gto., M\'exico}\\
}

\begin{document}

\maketitle

\begin{abstract}
Following a letter by Bassett, we show first that it is possible to find an 
analytical approximation to the error function in terms of a finite series of 
hyperbolic tangents from the supersymmetric (SUSY) solution of the 
P\"oschl-Teller eigenvalue problem in quantum mechanics (QM).  Afterwards, we 
show that the second order differential equation for the derivatives of Dawson's 
function can be found in another SUSY related eigenvalue problem, where the 
factorization of the simple harmonic oscillator Hamiltonian renders the 
wrong-sign Hermite differential equation, and that Dawson's second order 
differential equation possess a singular SUSY type relation to this equation.
\end{abstract}

\noindent {\bf Keywords:} Supersymmetry, error function, Dawson's function

\noindent {\bf PACS numbers:} 02.30.Gp, 02.30.Mv, 05.10.Ln, 11.30.Pb

\vspace*{3mm
}

The error function,\cite{ferr} which is defined by the integral
\begin{equation}
 \mbox{erf}(x) = \frac{2}{\sqrt{\pi}} \int_0^x e^{-t^2} dt \ , \label{erfun}
\end{equation}
is one of the most recurrent functions found in mathematical and physical sciences.  Nonetheless, it lacks an analytical expression in terms of elementary functions.

In 1996, Bassett proposed an analytical approximation to the error function in terms of the hyperbolic tangent,\cite{basset} 
\begin{equation}
 \mbox{erf}(x) \simeq \tanh \left( \frac{2}{\sqrt{\pi}}\, x \right) \ , \label{baprox}
\end{equation}
which converges asymptotically to the error function as $x \to\infty$.

Although other relations may render a closer approximation to the error 
function,\cite{otro1,otro2,otro3} it is possible that Bassett's be the simplest analytical approximation found in the literature allowing an analytical random number generator for the Gaussian function.

In this letter we present an approximation to the error function in terms of a finite series of hyperbolic tangents, where the accuracy is determined by the series length, which is inspired in a SUSYQM problem. 
We then consider Dawson's function, which is intimately related to the error function, and show that we can find a singular SUSY relation between the derivatives of Dawson's function and the eigenfunctions of the wrong-sign Hermite differential equation.
We finish the letter by proposing another approximation to Dawson's function in terms of a sum of Gaussians.


\section{The kink solution and the P\"oschl-Teller solutions}
\label{sectionIH}

In his article, Basset's approach was based on the fact that the hyperbolic tangent, which is found as the topological solution to the 1+1 dimensional partial differential equation for the soliton
\begin{equation}
 \phi_{tt}-\phi_{xx}=2 b^2 \left( \phi-\frac{1}{A^2}\phi^3 \right) \, ,
 \label{kink}
\end{equation}
has the same kink form than the error function.  However, since eq.(\ref{kink}) is a non-linear equation, it does not give any further guide on how to improve the accuracy of the approximation.  Therefore, Bassett proposed to extend his approximation using two hyperbolic tangents, as
\begin{equation}
 \mbox{erf}(x)\simeq \tanh\left( \frac{2}{\sqrt{\pi}}\, x \right)+ \frac{d~}{dx}
 \left( \alpha\, \tanh^p\left(x\right) \right) \ ,
 \label{bass2}
\end{equation}
where $\alpha$ and $p$ are parameters to be estimated.  

In order to find a succession of functions, where each one becomes an improved approximation to the error function, we propose to look at another physical problem, that of a quantum particle subject to a one dimensional P\"oschl-Teller potential.

The time independent Schrödinger equation for a particle in the presence of a modified P\"oschl-Teller potential \cite{PTP} is
\begin{equation}
 \left(-\frac{\hbar^2}{2m}\frac{d^2}{dx^2}-\frac{\alpha^2\lambda(\lambda+1)}{\cosh^2\alpha x}\right)\psi=E\,\psi \ ,
 \label{eq1}
\end{equation}
where $\alpha>0$ and the integer $\lambda>0$.  Infeld and Hull \cite{infeld} solved the factorization of the Hamiltonian of this problem: using dimensionless variables, with $\frac{\hbar^2}{2m}=1$, the factorization is realized in terms of the raising and lowering operators
\begin{equation}
 ^\pm H^\lambda_\ell=\frac{1}{\alpha \sqrt{\lambda^2 - \ell^2}}\left( \alpha \lambda\tanh (\alpha x) \pm\frac{d}{dx} \right) \ ,
\end{equation}
where the plus/minus sign is for the lowering/raising operator of the $\lambda$ parameter.

For this problem, the energy eigenvalues are
\begin{equation}
 E_n=-\alpha^2(\lambda-n)^2,    \hspace{3cm} n=0,1,2...<\lambda \ ,
\end{equation}
where $n=\lambda-\ell$, and the wave functions $ \psi_n(x)\equiv\psi_{\lambda-n}^\lambda(x)=\psi_\ell^\lambda(x) $
are found by successive application of the raising operator $ ^- H^\lambda_\ell$ on the normalized functions
\begin{equation}\label{psi0}
 \psi_\ell^\ell(x)=\sqrt{\frac{\alpha\Gamma(\ell+\frac{1}{2})}{\sqrt{\pi}\Gamma(\ell)}}\cosh^{-\ell}\alpha x \ .
\end{equation}
For given $\lambda$ and $\alpha$, the ground function $\psi_0$ is given by this function, with $\ell=\lambda$.


\subsection{Finite series approximation}

In eq.(\ref{psi0}), $\sech^{\lambda}(\alpha x)$ already has the bell shape of the Gaussian function for $\lambda=1$, and the resemblance improves as $\lambda$ increases.  Since any even power of $\sech(\alpha x)$ is easily integrated into a series of hyperbolic tangents, we propose to define the function 
$\mbox{erf}_\lambda(x)$ as
\begin{equation}\label{aerf}
\mbox{erf}_\lambda(x)\equiv \frac{2\alpha\Gamma(\lambda+\frac{1}{2})}{\sqrt{\pi}\, \Gamma(\lambda)}
\int_0^x\sech^{2\lambda} (\alpha y) \, dy \ ,
\end{equation}
where $\lambda=$1,2,3,..., which has the asymptotic values
\begin{equation}\label{asimp}
\mbox{erf}_\lambda(\pm \infty)= \pm 1 \, .
\end{equation}
A plot of this function for increasing values of the parameter $\lambda$ is shown on the left panel of Fig.1, together with the error function. On the right panel we plot the difference between the error function and the different forms of erf$_\lambda$($x$), showing that as 
$\lambda$ increases, the difference with respect to the error function decreases.

\begin{figure}[H]\begin{center}
\subfigure{\includegraphics[width=0.4\textwidth]{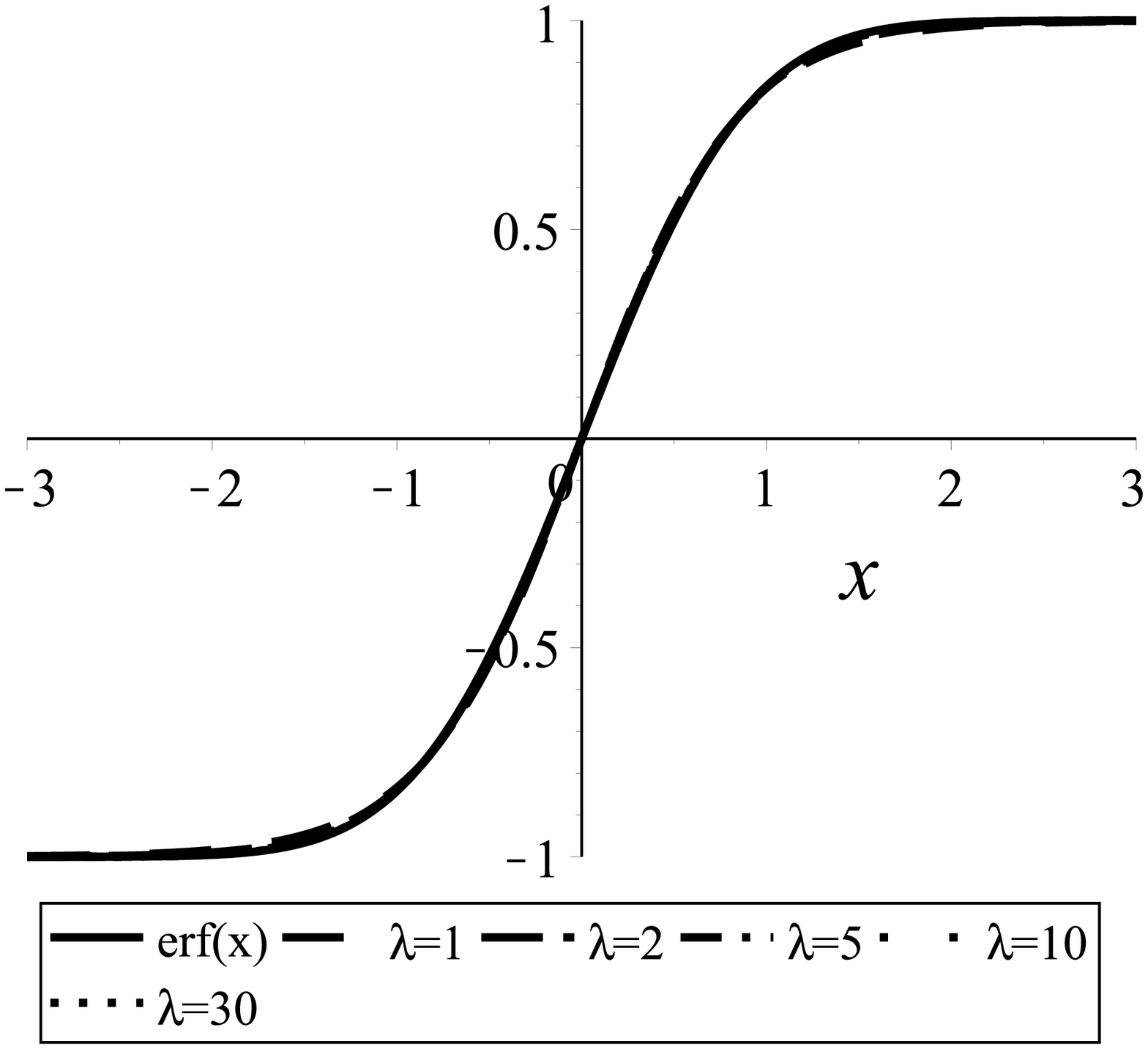}}
\hspace{3mm}
\subfigure{\includegraphics[width=0.54\textwidth]{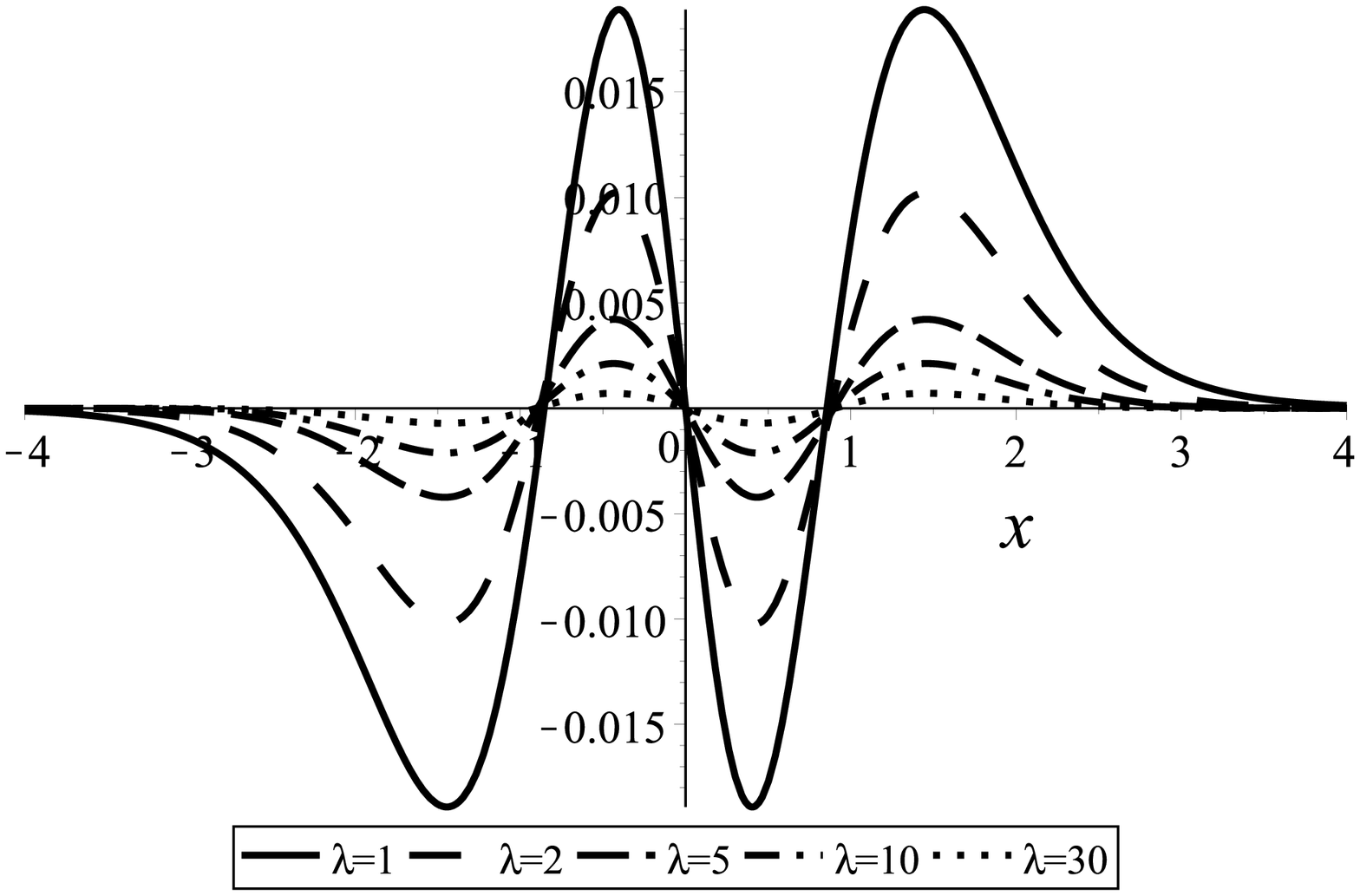}}
\caption{\label{erff}
Left panel: erf($x$) together with erf$_\lambda$($x$) for increasing values of the parameter $\lambda$.
Right panel: the difference erf($x$)$-$erf$_\lambda$($x$) for increasing values of the parameter $\lambda$.  As explained in the text, the parameter $\alpha$ has been chosen to make the two peaks of the largest differences of similar height.}
\end{center}\end{figure}

For integer $\lambda$, the integral in eq.(\ref{aerf}) is easily evaluated, leading to the series
\begin{equation}\label{aerfser}
\mbox{erf}_\lambda(x) =
\frac{2\, \Gamma(\lambda+\frac{1}{2})}{\sqrt{\pi}\, \Gamma(\lambda)}
\sum_{k=0}^{\lambda-1} \left( \!\!\! \begin{array} {c} \lambda-1 \\ k \end{array} \!\!\! \right)   
\frac{(-1)^{k}}{2k+1} \, \tanh^{2k+1}(\alpha x) \ .
\end{equation}
It is interesting to note here that this series is used to find the range of validity of the parameter that determines the SUSY partner potentials for the P\"oschl-Teller problem in QM \cite{rosas}.
Since the series (\ref{aerfser}) is defined for all $x$, we have defined an analytical finite series to represent the error function.

Now, the parameter $\alpha$ only appears in the argument of $\tanh$ in 
eq.(\ref{aerfser}), and does not interfere with the asymptotic values in 
eq.(\ref{asimp}); then, it is possible to use it here as a fine tunning parameter to minimize the maximum  difference between the error function and erf$_\lambda$($x$) for all $x$, as can be seen in the right panel of Fig.\ref{erff}.
With this approach, the maximum difference between the error function and erf$_\lambda(x)$ is less than 0.0022 for $\lambda=10$ and less than 0.00072 for $\lambda=30$.

\begin{figure}[H]\begin{center}
\subfigure{\includegraphics[width=0.55\textwidth]{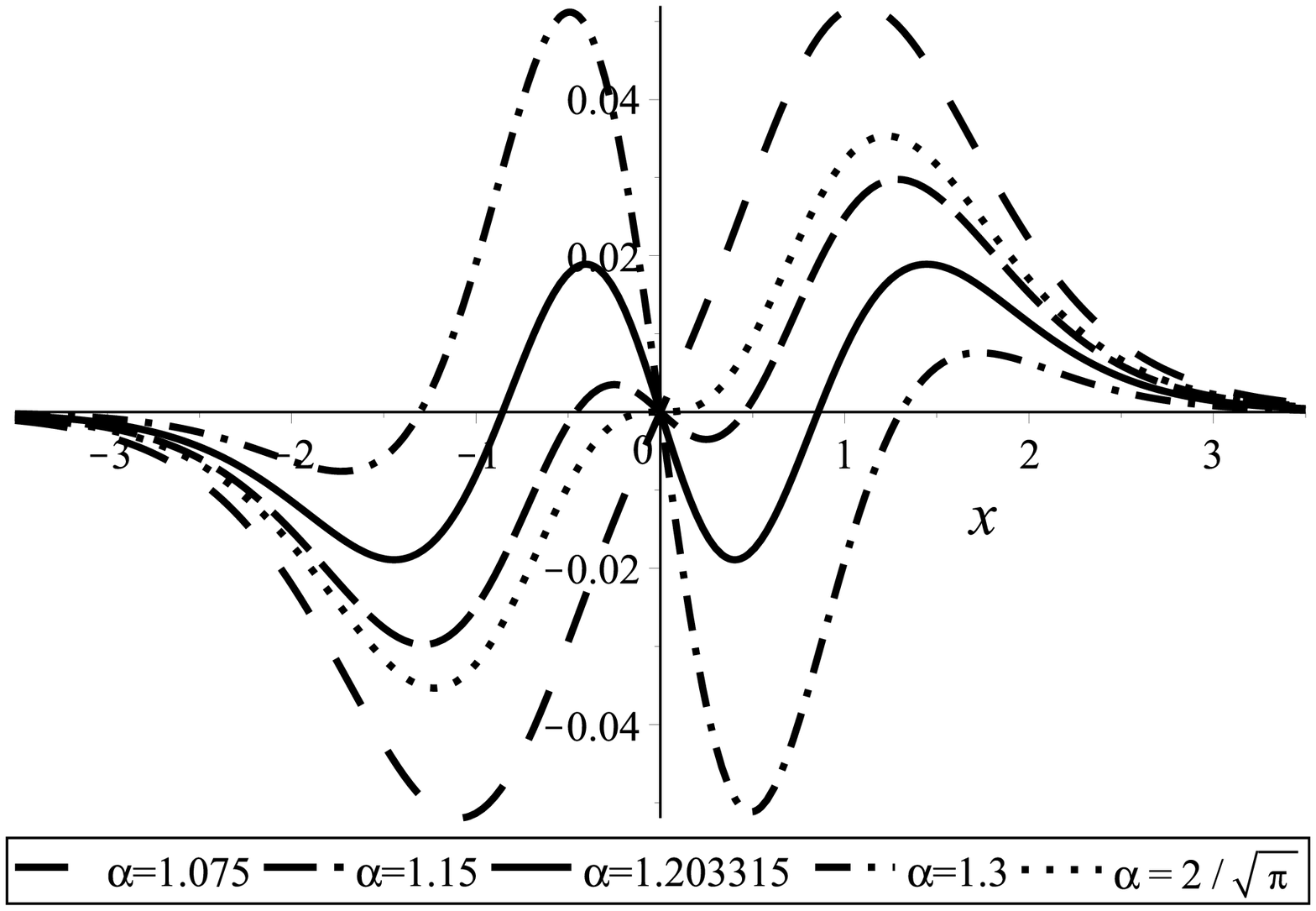}}
\subfigure{\includegraphics[width=0.42\textwidth]{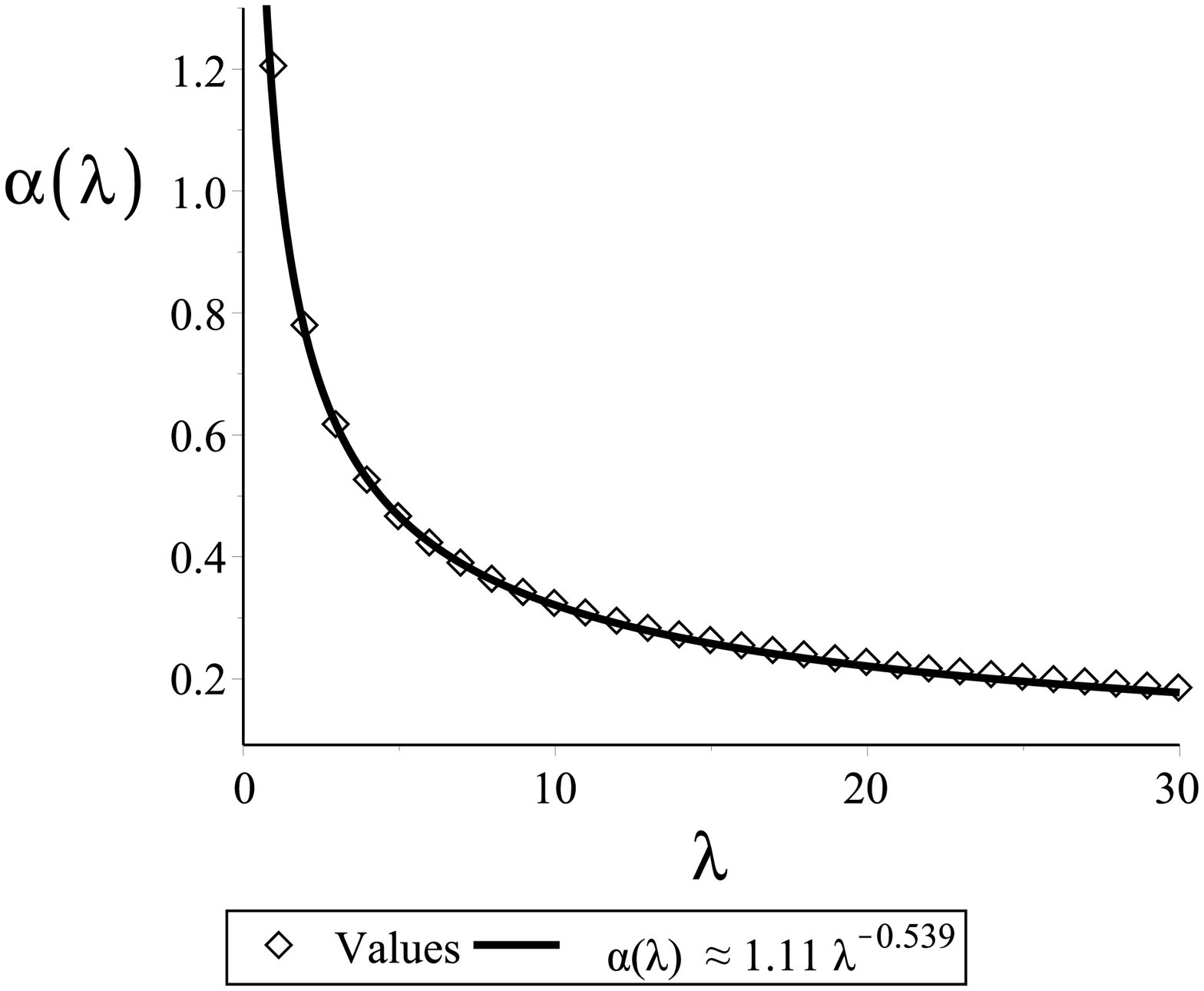}}
\caption{\label{alfa}
Left panel: The difference erf$(x)-$erf$_1(x)$ for $\alpha=$1.075, 1.15, and 1.3 (dashed curves.)  Our choice of $\alpha=1.203315$ is plotted with a solid line, while Basset's case of $\alpha=2/\sqrt{\pi}$ is plotted using a dotted line.
Right panel: the values of $\lambda$ and $\alpha$ from Table \ref{tabla1}, in markers; the curve is the function $1.11 \lambda^{-0.539}$.}
\end{center}\end{figure}

\begin{table}[h]
\caption{The values of $\alpha$ that minimize the difference erf$(x)-$erf$_\lambda(x)$, for different values of $\lambda$.}
\begin{center}
\begin{tabular}{ c c } 
 \hline\hline
 $\lambda$ & $\alpha$ \\  \hline
 1 & 1.203315 \\ 
 2 & 0.778004 \\ 
 3 & 0.615589 \\ 
 4 & 0.524697 \\ 
 5 & 0.464819 \\ 
 7 & 0.388543 \\ 
 10 & 0.3224 \\ 
 15 & 0.261549 \\ 
 20 & 0.225779 \\ 
 30 & 0.183755 \\ 
\hline
\end{tabular}
\label{tabla1}
\end{center}
\end{table}

In the left panel of Fig.\ref{alfa} we show how $\alpha$ modules the difference erf($x)$\-erf$_\lambda(x)$ for the case $\lambda=1$, using only one hyperbolic tangent.  The dashed curve corresponds to $\alpha=1.075$, and the long-dashed and dash-dotted lines are for $\alpha$=1.15 and 1.3, respectively.  Our choice, using the approach explained above, has 
$\alpha=1.203315$ and is plotted with a solid line, while Bassett's case of $\alpha=2/\sqrt{\pi}$ is plotted using a dotted line.  In this case, our choice renders a maximum error which is about half of the maximum error in Bassett's approximation.  

In Table \ref{tabla1} we give the values of $\alpha$ which minimize the difference between the error function and erf$_\lambda(x)$ for different values of $\lambda$.  As it is shown in the right panel of Fig.(\ref{alfa}), the $\alpha$ dependence on $\lambda$ almost follows the function 1/$\sqrt{\lambda}$, but the exact dependence is difficult to find since it involves transcendental functions.

\subsection{Gaussian distribution generator}

One immediate application when having an analytical approximation to the error function is that one can define a generator for the Gaussian function.  If $f(x)$ represents a probability density function (p.d.f.) normalized to unity, defined for $-\infty<x<\infty$, and $F(a)$ is the corresponding cumulative distribution function, $F(a)=\int_{-\infty}^a f(x)dx$, where the inverse of $F(x)$ exists, then if $u$ is uniformly generated in (0,1), we can find a unique $x$ chosen from the p.d.f. by assigning $x=F^{-1}(u)$.

\begin{figure}[H]\begin{center}
\subfigure{\includegraphics[width=0.3\textwidth]{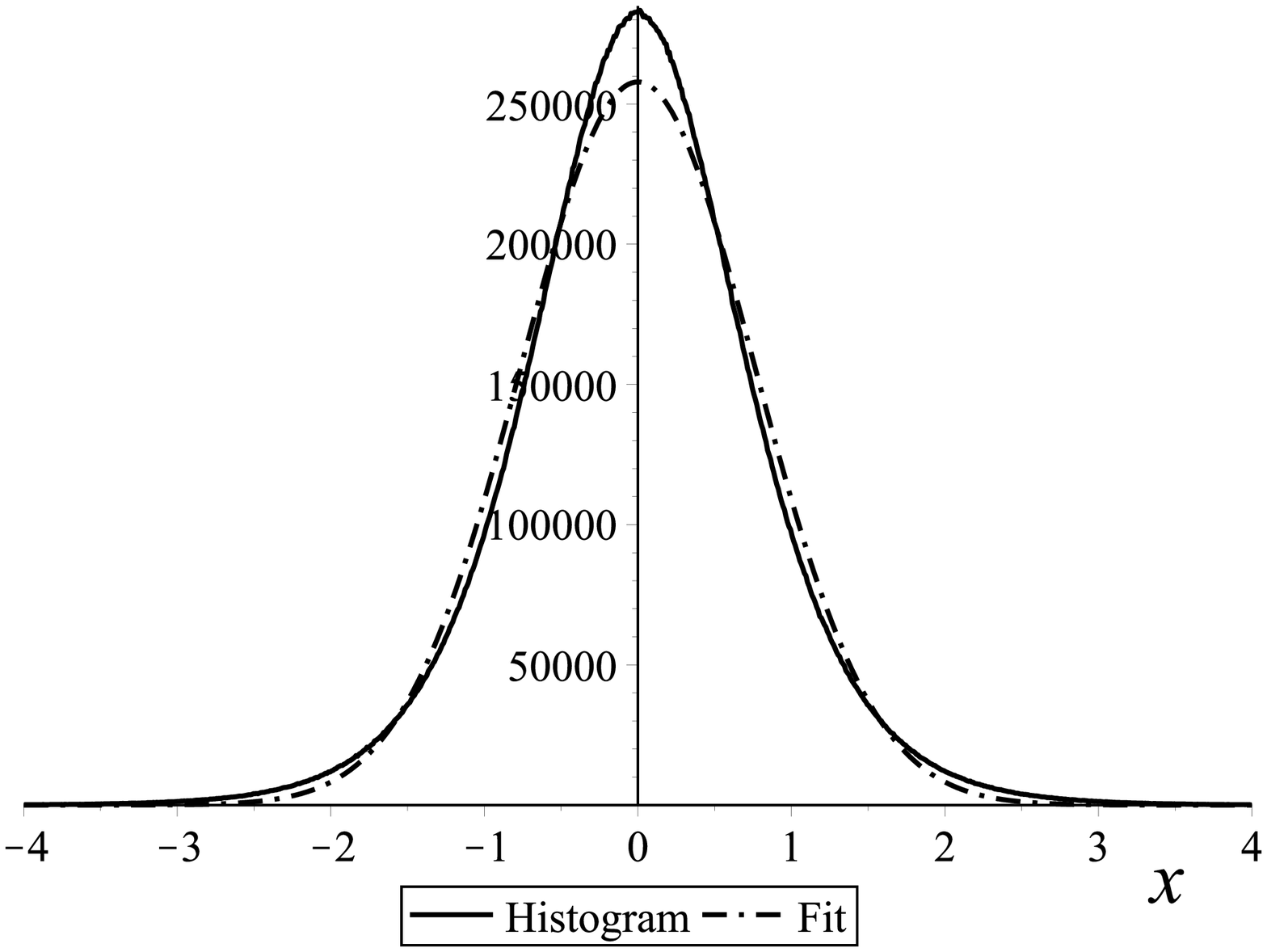}}
\subfigure{\includegraphics[width=0.3\textwidth]{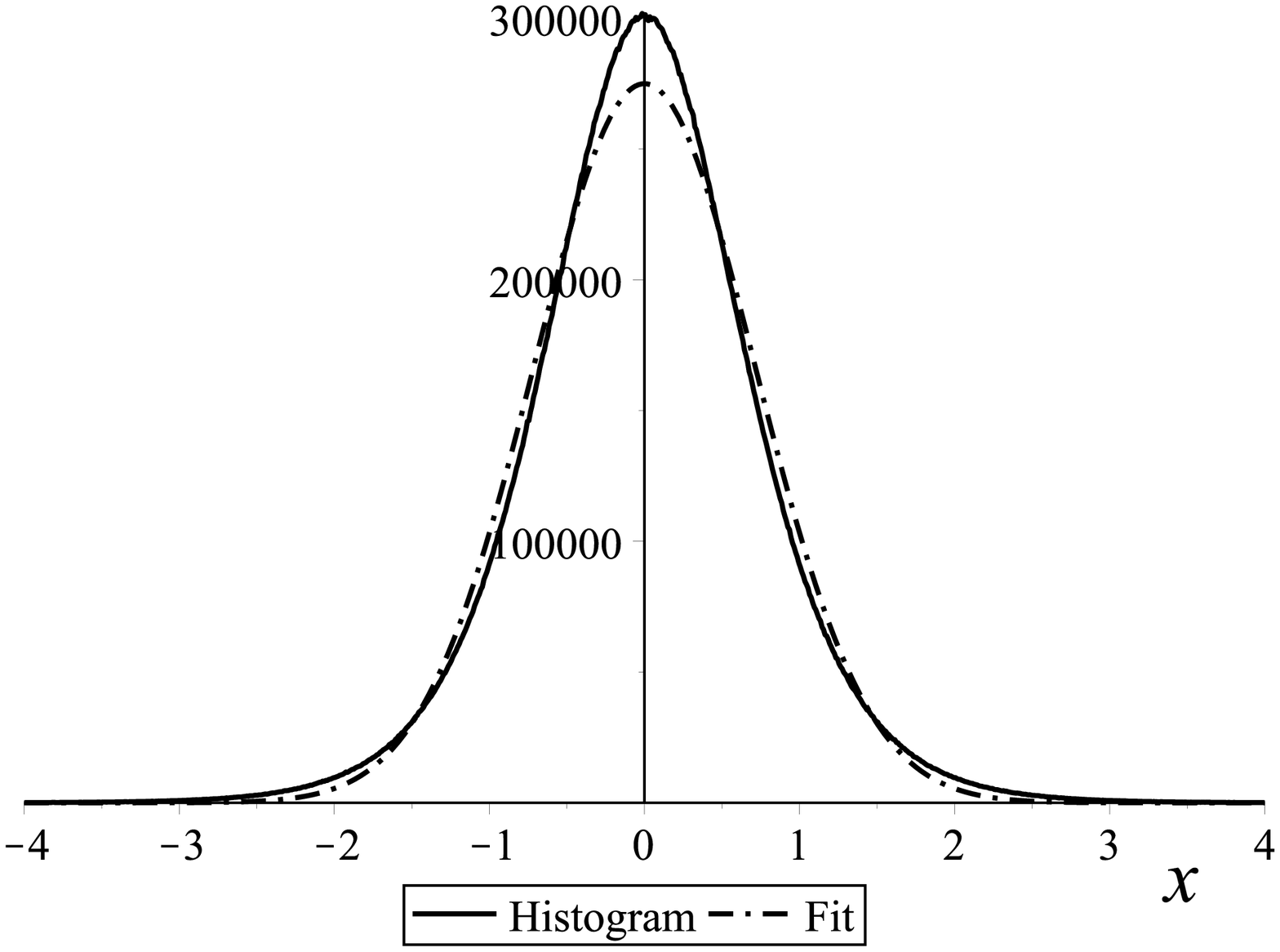} }
\subfigure{\includegraphics[width=0.3\textwidth]{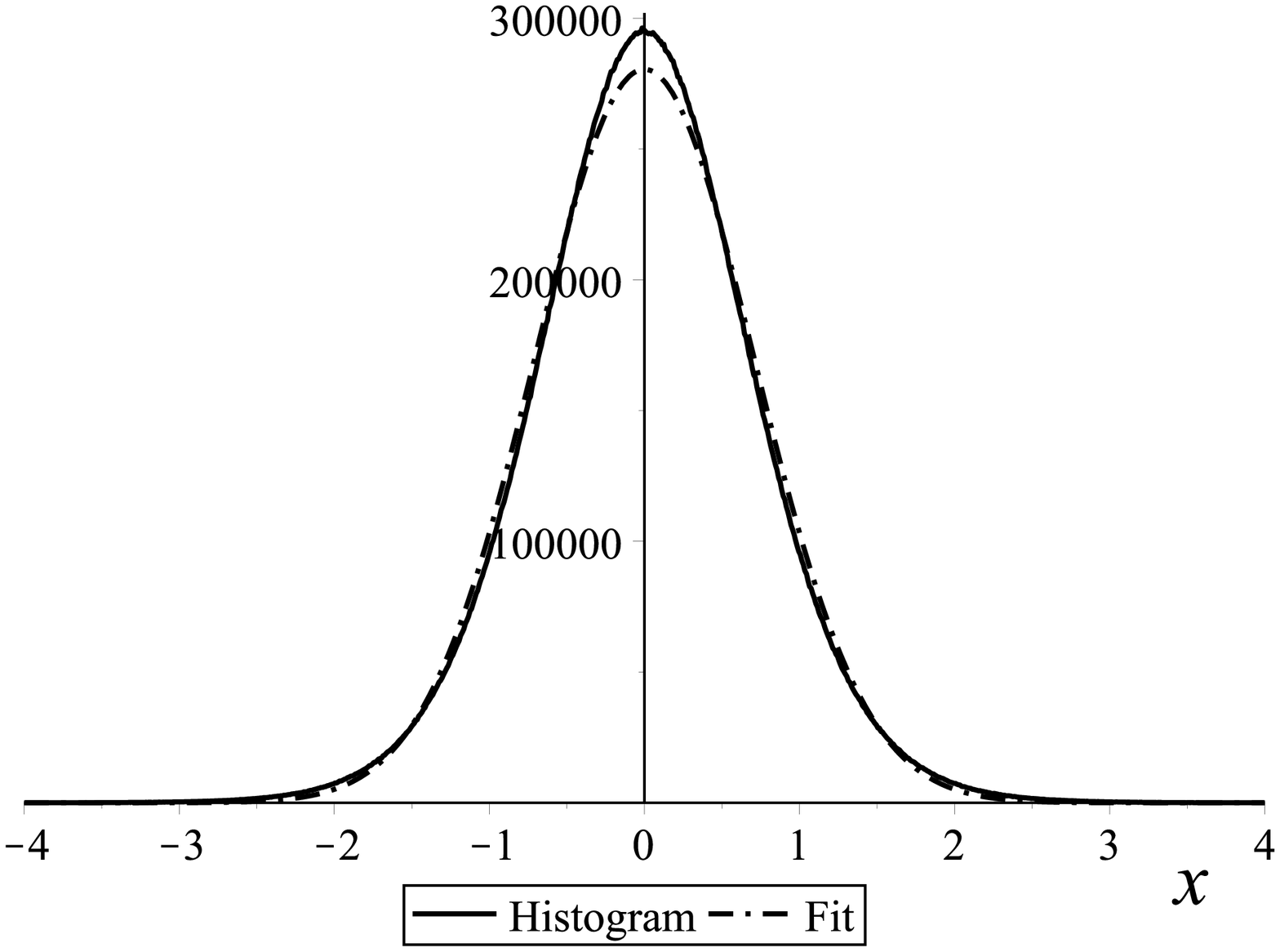} }
\caption{\label{gaussgen}
Histograms (solid lines) and fits (dashed lines) to events generated using Bassett's approximation, eq.(2) (left panel), and erf$_1(x)$ with $\alpha=1.203315$ (central panel), and erf$_2(x)$ with $\alpha=0.7865$. The means and widths obtained were $\mu=0.1477\times 10^{-3}$, $\sigma=0.7618$, $\mu=-0.2008\times 10^{-3}$, $\sigma=0.7143$, and $\mu=-0.0227\times 10^{-3}$, $\sigma=0.7071$, respectively.
}
\end{center}\end{figure}

Here we compare the p.d.f.'s obtained when we generated fifty million events using Bassett's form for the error function given by eq.(2), and our definition eq.(11) for the cases $\lambda=1$, with $\alpha=1.203315$, and $\lambda=2$, with $\alpha=0.7865$.  
The histograms of randomly generated numbers were fitted using a Gaussian function.  If the generator had been exact the width of the Gaussian would be $1/\sqrt{2}$. On the left panel of Fig.(\ref{gaussgen}) we show the fit to events generated using Bassett's approximation.  On the central and right panel we show the histograms and fits to events generated using eq.(11) with $\lambda=1,2$.  The widths obtained in the three cases were 0.7618, 0.7143 and 0.7071, respectively.  In the case $\lambda=2$, $\alpha$ was chosen to approximate the correct width.


\section{Dawson's function}

As it is well known, Dawson's function is directly related to the error function of imaginary argument \cite{daw1}
\begin{equation}\label{dawf}
 F(x) = e^{-x^2} \int_0^x e^{y^2} dy =-\frac{\sqrt{\pi}\, i}{2}\: e^{-x^2} \mbox{erf}(i x)
\end{equation}
and its properties are better determined in terms of its derivatives,
\begin{subequations}
\begin{align}\label{der1}
 F'(x) & = -2 x F(x) +1  \, , \\ \label{derkm1}
 F^{(k+1)}(x) + 2x & F^{(k)}(x) + 2k F^{(k-1)}(x) =0 \, .
\end{align} \label{dder} 
\end{subequations}

Attempts have been made to find a simple representation of this function in terms of elementary functions, for example, that of Cody {\it et al.},\cite{cody} who proposed a representation of $F(x)$ in terms of rational approximations,
\begin{subequations}
\begin{align}\label{uno}
F_{lm}(x) &= x R_{lm}(x^2) \, , & |x|\leq 2.5 \\ \label{dos}
F_{lm}(x) &= \frac{1}{x} R_{lm}\! \left( \frac{1}{x^2} \right) \, , & 2.5 \leq |x|\leq 3.5 \ ; \ 3.5 \leq |x|\leq 5.0 \\ \label{ters}
F_{lm}(x) &= \frac{1}{2x} \left[ 1+\frac{1}{x^2} R_{lm}\left( \frac{1}{x^2} \right) \right] \, , & 5.0 \leq |x|
\end{align} \label{codyf} 
\end{subequations}
where $R_{lm}(x)$ are rational functions of degree $l$ in the numerator and $m$ in the denominator.

Obviously, it is not possible to extend our finite series approximation of the error function to Dawson's function, and we shall need to use a different approach to find an analytical approximation.  But before doing so, we shall show here that there exists 
a singular SUSY type relation between Dawson's function and its derivatives, and the eigenfunctions of the wrong-sign Hermite differential equation, which is found in a SUSY type factorization of the Hamiltonian of the simple harmonic oscillator 
(SHO).\cite{rafart}


\subsection{Generalized factorization of the SHO Hamiltonian}

In a previous article \cite{rafart} we have developed two factorizations of the SHO Hamiltonian in terms of two non-selfadjoint operators
\begin{subequations}
\begin{align}\label{bmenos}
 B^- &= \frac{1}{\sqrt{2}}\left(\alpha^{-1}(x)\frac{d}{dx}+\beta(x)\right),\\ \label{bmas}
B^+ &= \frac{1}{\sqrt{2}}\left(-\alpha(x)\frac{d}{dx}+\beta(x)\right).
\end{align} \label{Bs}
\end{subequations}
In the first factorization, we required that $B^-B^+=H+\frac{1}{2}$.  Upon inverting the product, $B^+B^-$, the two parameter solutions for $\alpha(x)$ and $\beta(x)$ defined a Sturm-Liouville equation which included the quantum mechanics SHO equation, its SUSY partners,\cite{mielnik} and Hermite's equation, as particular cases for defined regions of the two-parameter space.\cite{rafart}

In a second factorization, we proposed that the Hamiltonian be factorized as $B^+B^-=H-\frac{1}{2}$, which is possible if now the functions $\alpha(x)$ and 
$\beta(x)$ depend on a single parameter, $\gamma_3$, and are given by
\begin{equation}
  \alpha_{\gamma_3}(x) = \sqrt{1+\gamma_3e^{x^2}}~,\qquad \beta_{\gamma_3}(x) = \frac{x}{\sqrt{1+\gamma_3e^{x^2}}}.
\end{equation}
The inverse operator product $B^-B^+$ now defines a new eigenvalue equation
\begin{equation}
 {\cal L}_{\gamma_3} \mbox{H}_{n}^{\gamma_3}+\lambda_n\omega_{\gamma_3}(x)\mbox{H}_{n}^{\gamma_3}=0,
\end{equation}
where
\begin{equation}
 {\cal L}_{\gamma_3} = \left( 1+\gamma_3e^{x^2} \right)\frac{d^2}{dx^2} + 2\gamma_3xe^{x^2}\:\frac{d}{dx}
+\frac{\gamma_3e^{x^2}+\gamma_3^2e^{2x^2}-x^2}{ 1+\gamma_3e^{x^2}}
\end{equation}
is a one-parameter self-adjoint operator with the weight function $\omega_{\gamma_3}(x) = 2 \left( 1+\gamma_3e^{x^2} \right)$, and it is isospectral to the quantum SHO Hamiltonian, which is obtained in the limit $\gamma_3\rightarrow 0$.  The eigenfunctions in this case are
\begin{equation}\label{ufunction}
\H_n^{\gamma_3}(x)=B^- \psi_{n+1}(x) \ ,
\end{equation}
where $\psi_{n}(x)$ are the SHO eigenfunctions.

For this work, it is very interesting to note that in the large limit $\gamma_3\gg 1$,
one can obtain the wrong-sign Hermite's differential equation
\begin{equation} \label{g-osc}
\left[ \frac{d^2~}{dx^2} + 2x \frac{d~}{dx} +2(n+1) \right] \widetilde H_n(x)=0~,
\end{equation}
which differs from the Hermite equation only in the sign in front of the first 
derivative. The corresponding eigenfunctions are of the quantum oscillator type, 
but vanishing faster due to a squared exponential factor, 
$\widetilde H_n(x) = e^{-x^2} H_n(x)$.  Also, from the Hermite polynomials' 
recursion relations it is easy to find the raising and lowering operators for 
these functions:
\begin{subequations}
\begin{align}\label{hnmenos}
\left( \frac{d~}{dx}+2x \right) \widetilde H_n(x) &= 2n \widetilde H_{n-1}(x),\\ 
\label{hnmas}
-\frac{d~}{dx} \widetilde H_n(x) &= \widetilde H_{n+1}(x).
\end{align} \label{tildehn}
\end{subequations}
Note that the reversed sign in the first derivative term of eq.(\ref{g-osc}) 
produces these “reversed" Hermite polynomials' recursion relations, where $n$-th 
eigenfunction is just the derivative of the previous one.


\subsection{Dawson's eigenfunctions, and a singular SUSY relation}

\begin{figure}[H]\begin{center}
\subfigure{\includegraphics[width=0.18\textwidth]{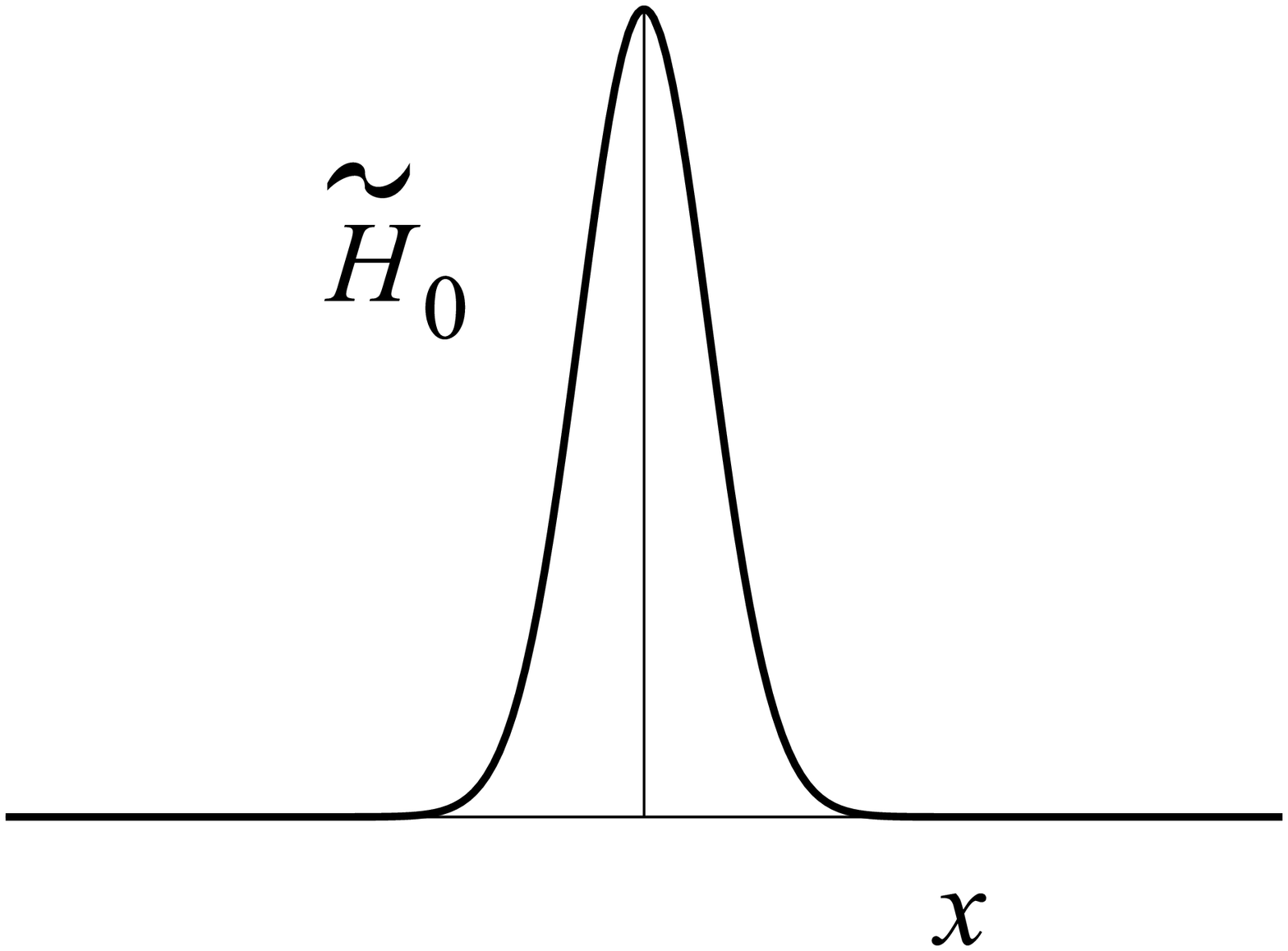}}
\subfigure{\includegraphics[width=0.18\textwidth]{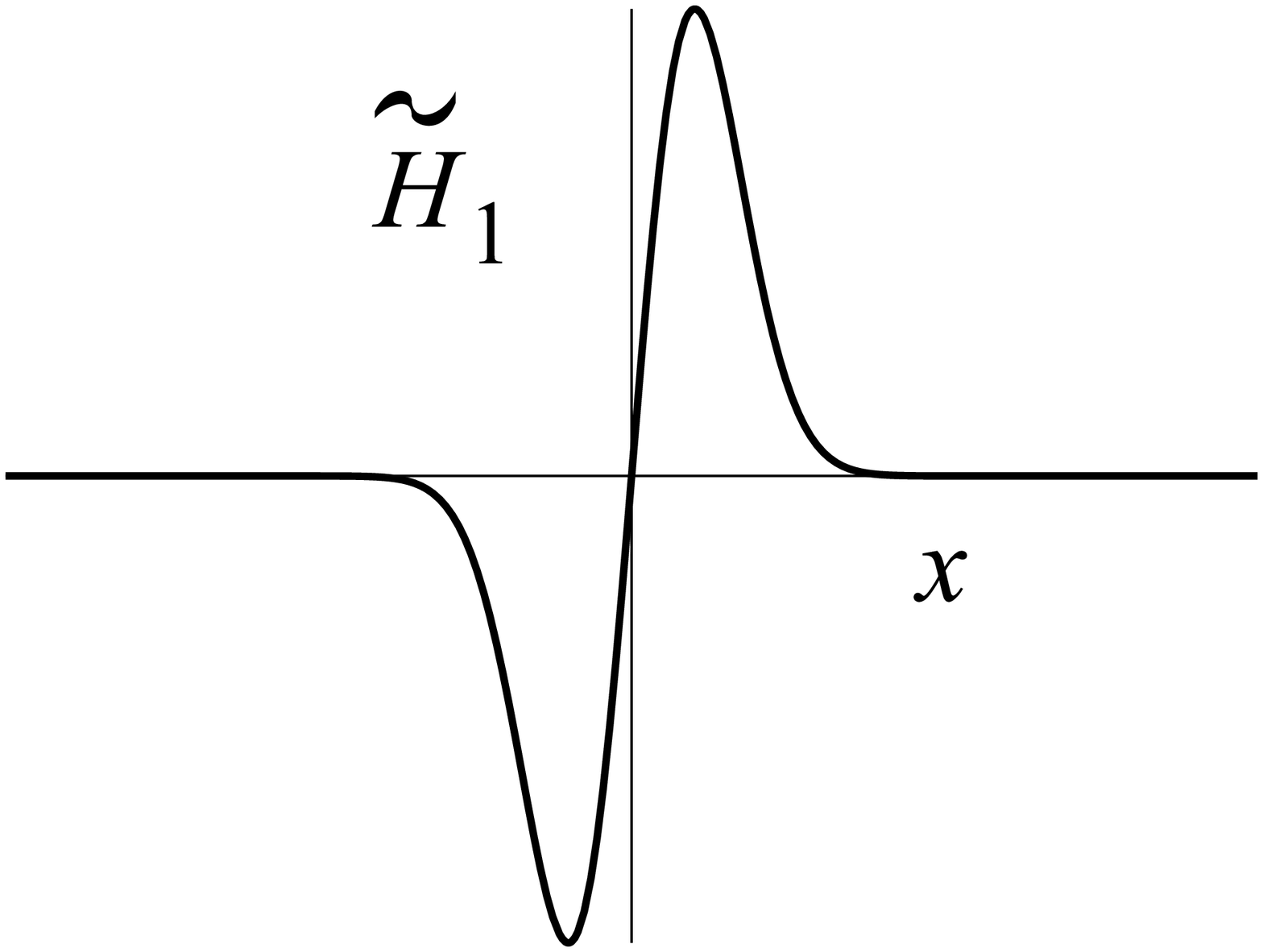}}
\subfigure{\includegraphics[width=0.18\textwidth]{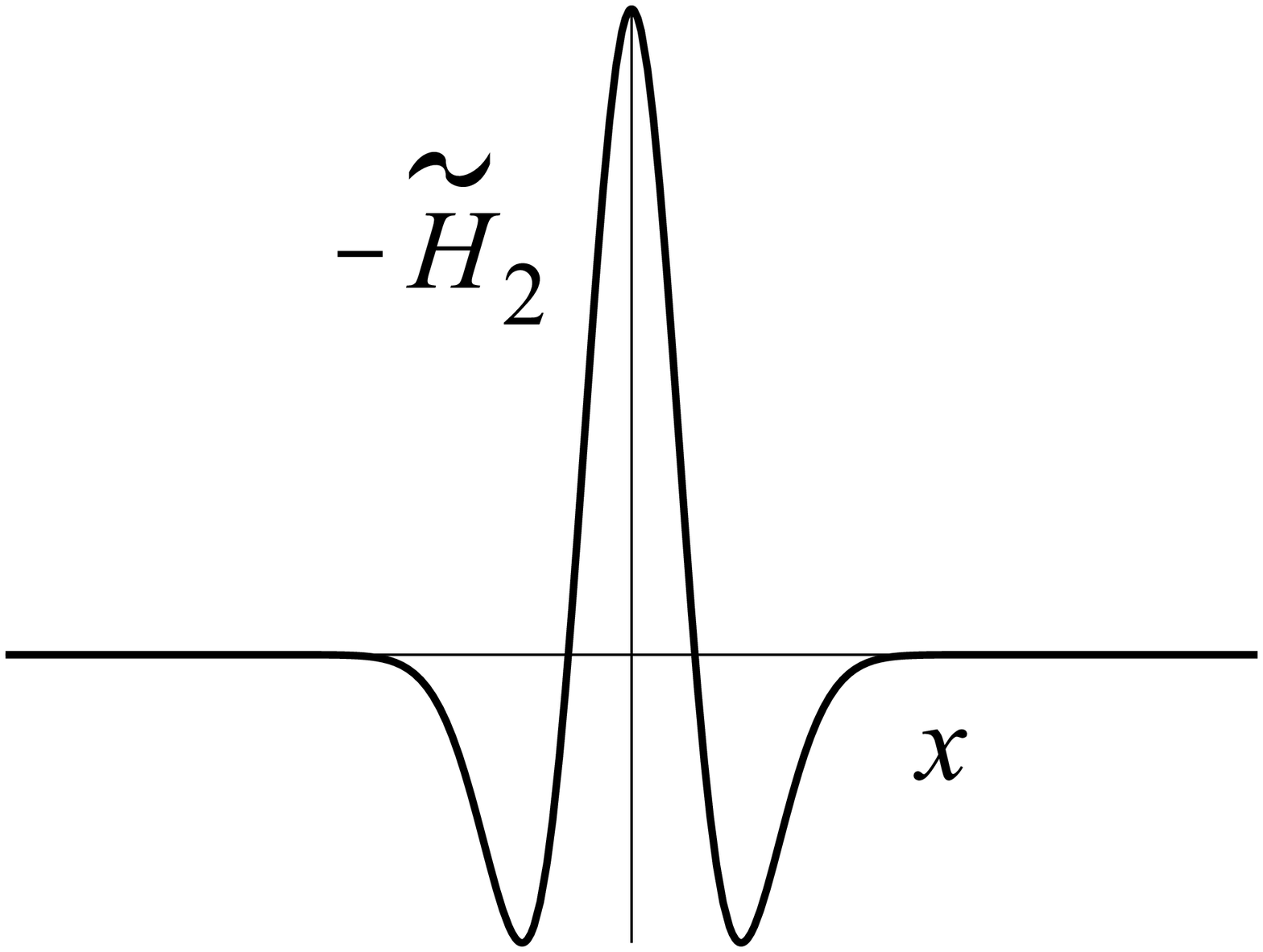}}
\subfigure{\includegraphics[width=0.18\textwidth]{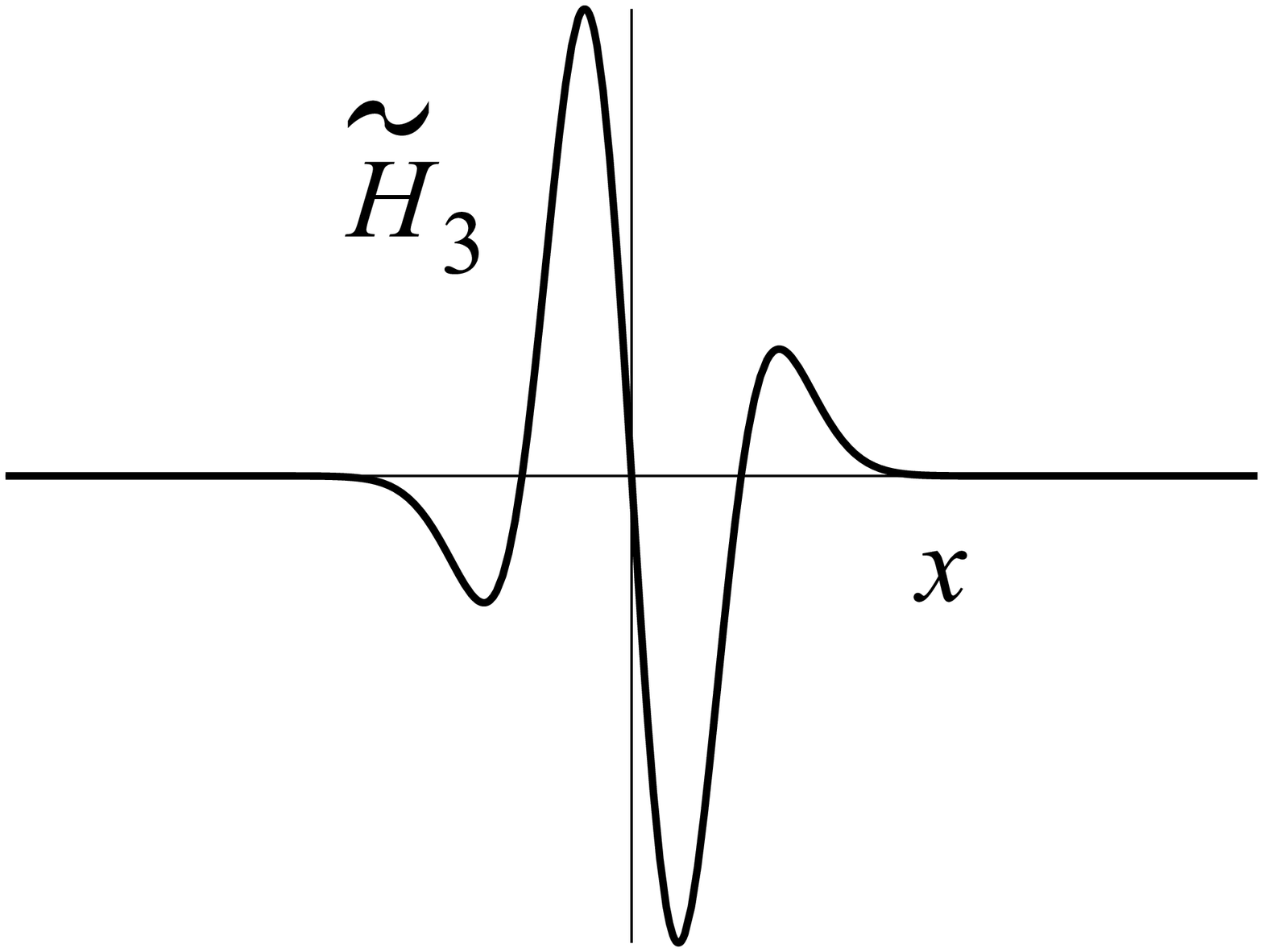}}
\subfigure{\includegraphics[width=0.18\textwidth]{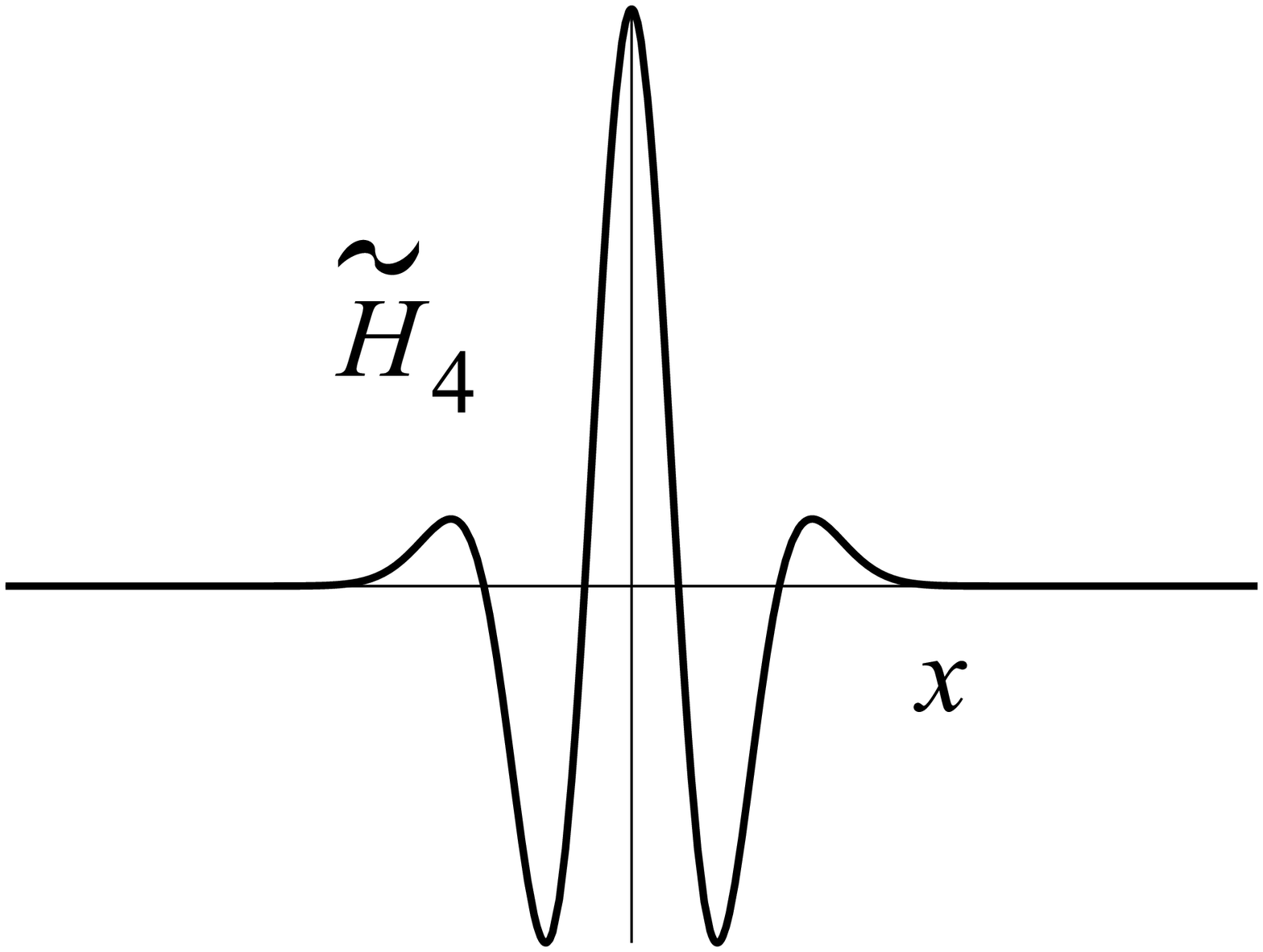}}\\
\vspace*{-7mm}
\subfigure{\includegraphics[width=0.18\textwidth]{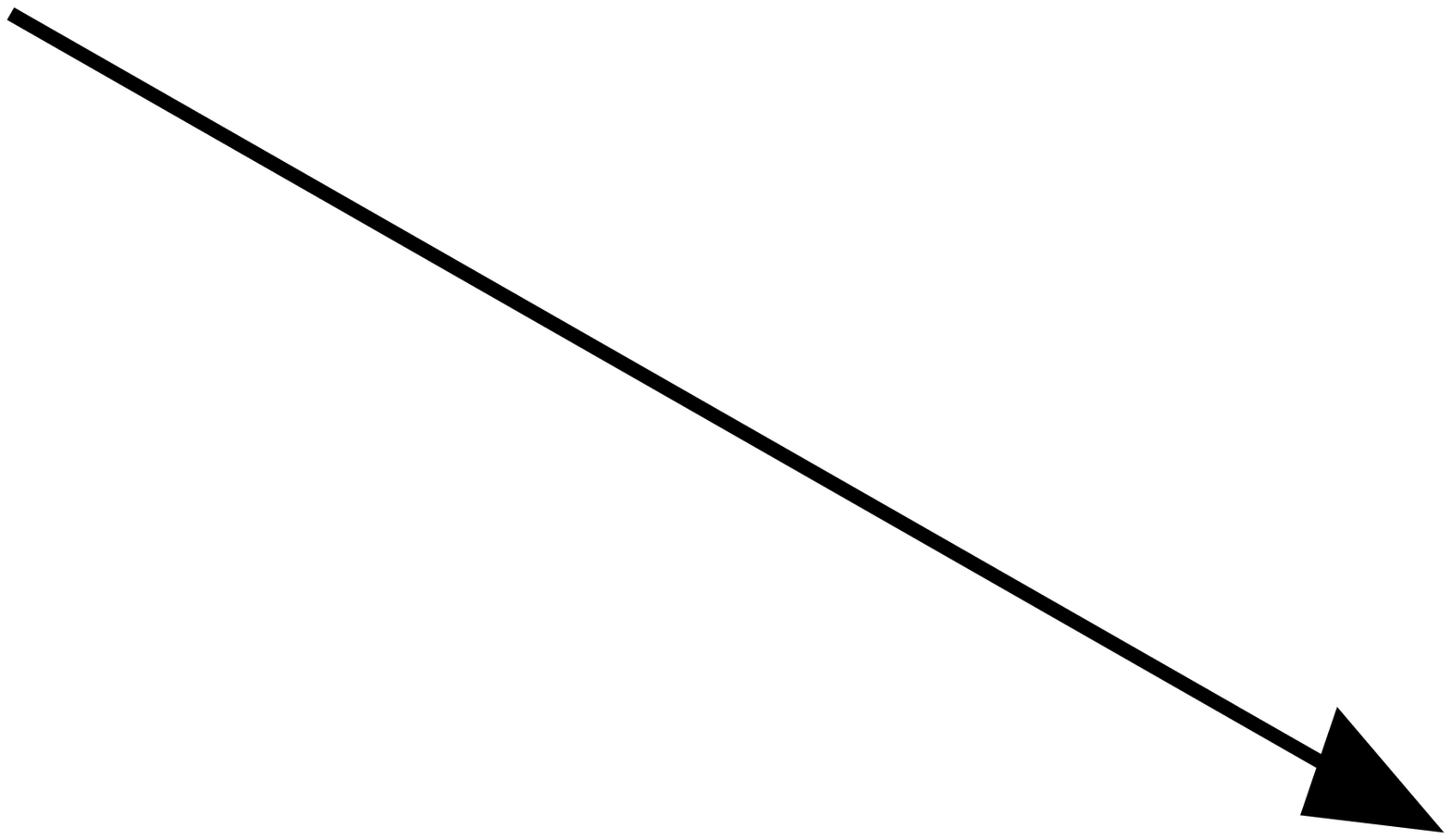}}
\subfigure{\includegraphics[width=0.18\textwidth]{figs-raf/arrow}}
\subfigure{\includegraphics[width=0.18\textwidth]{figs-raf/arrow}}
\subfigure{\includegraphics[width=0.18\textwidth]{figs-raf/arrow}}
\hspace*{2cm}\\
\vspace*{-8mm}
\subfigure{\includegraphics[width=0.18\textwidth]{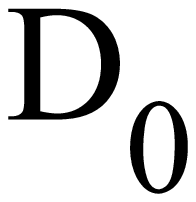}} 
\subfigure{\includegraphics[width=0.18\textwidth]{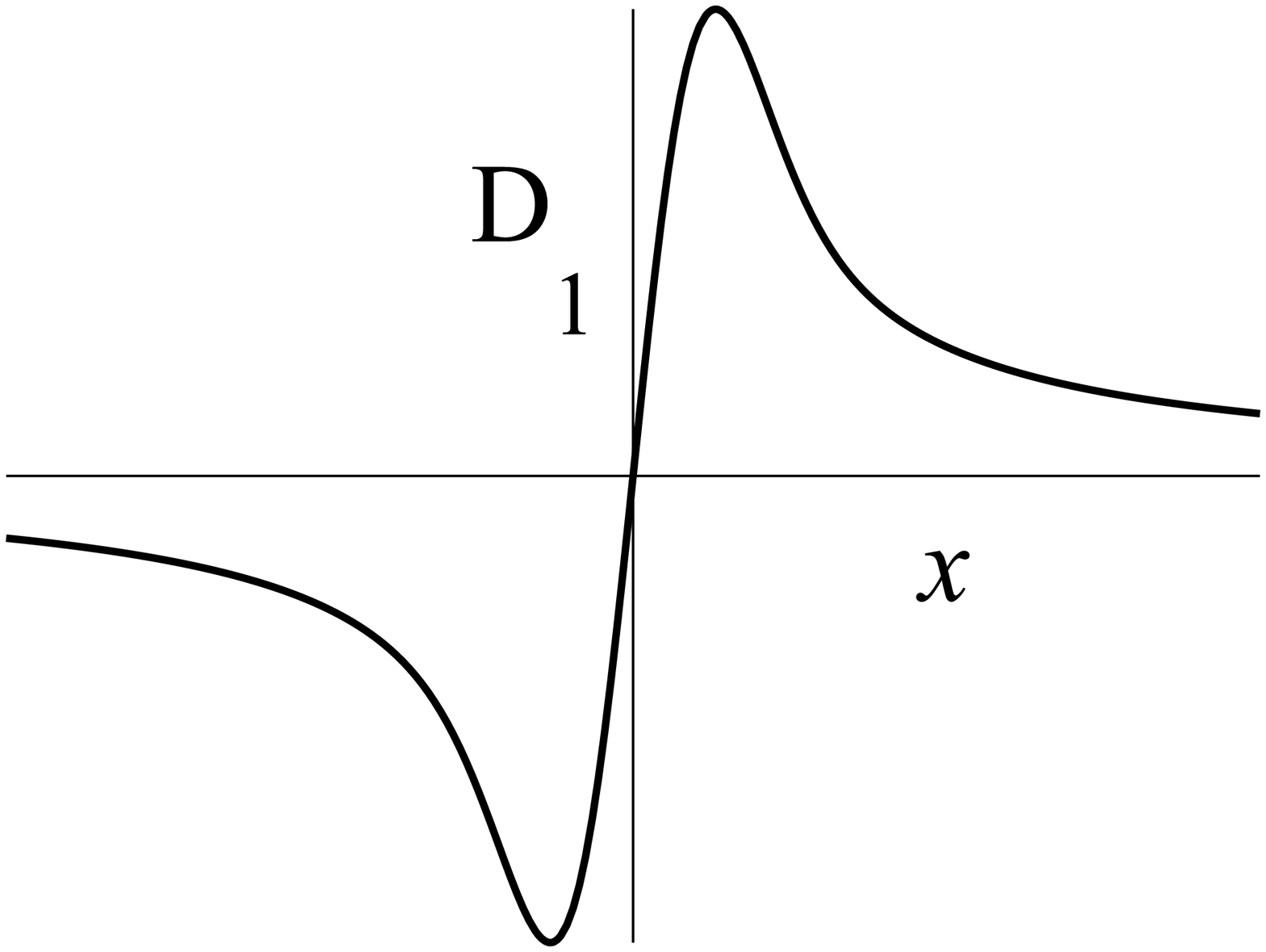}}
\subfigure{\includegraphics[width=0.18\textwidth]{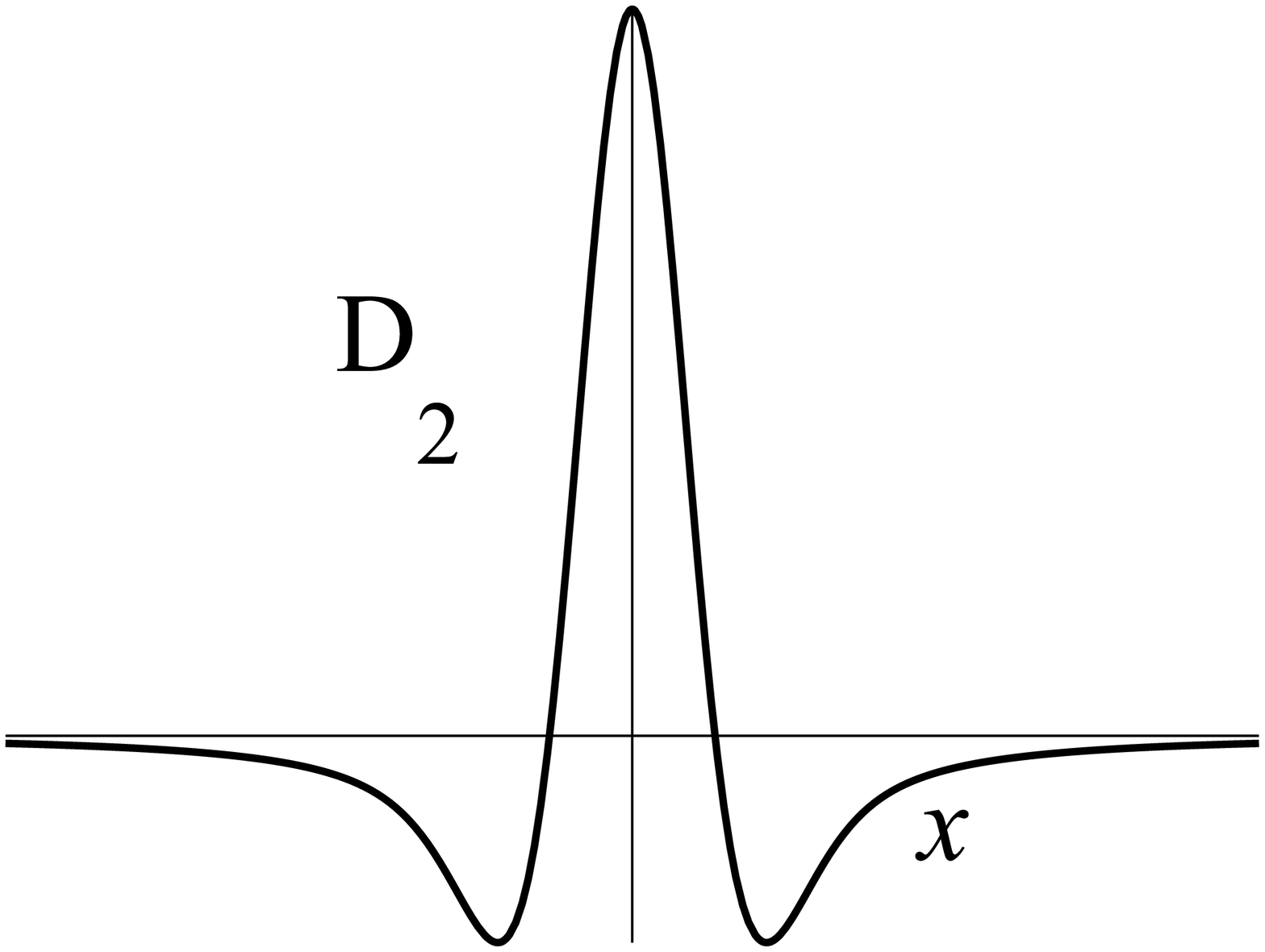}}
\subfigure{\includegraphics[width=0.18\textwidth]{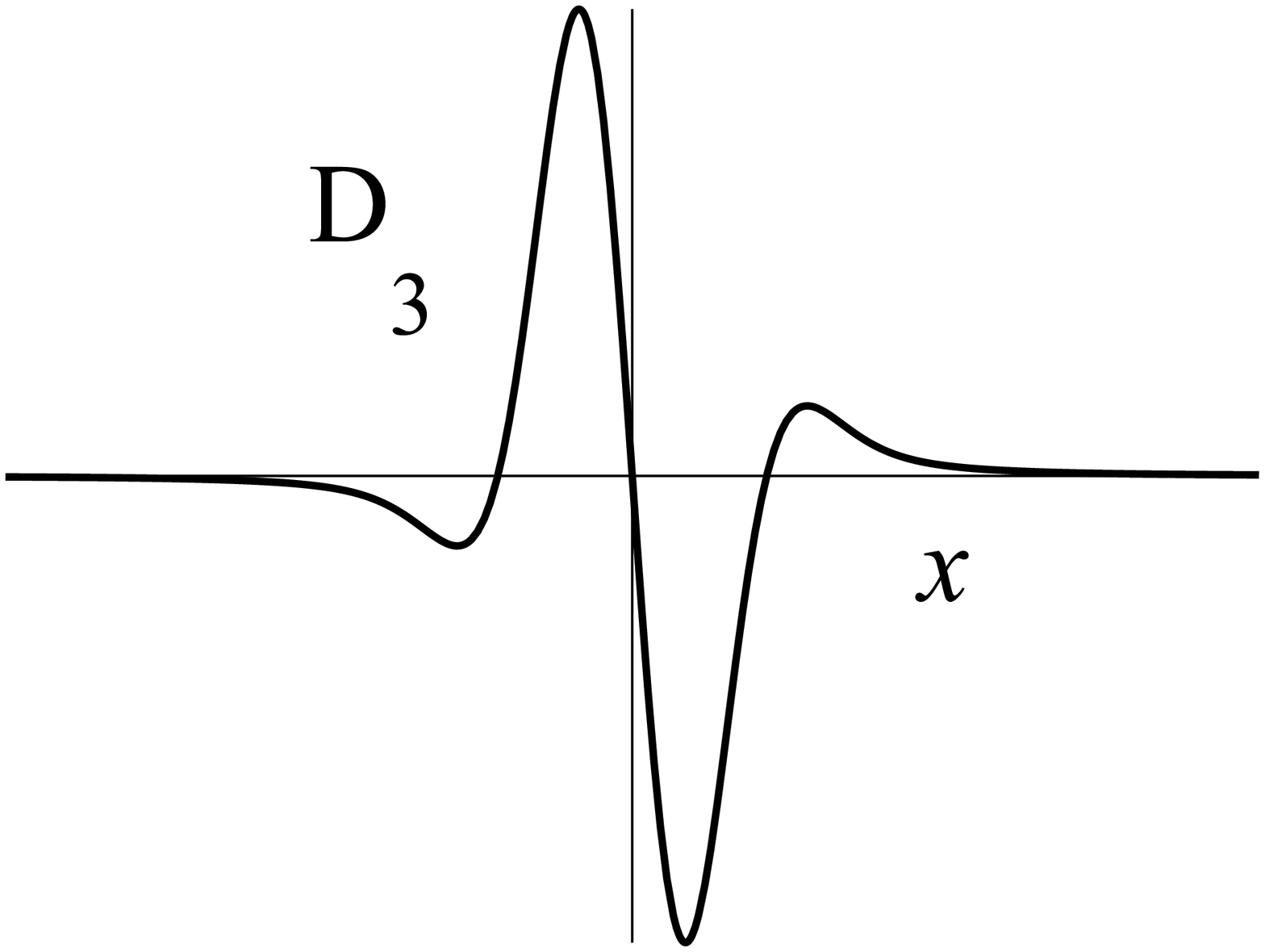}}
\subfigure{\includegraphics[width=0.18\textwidth]{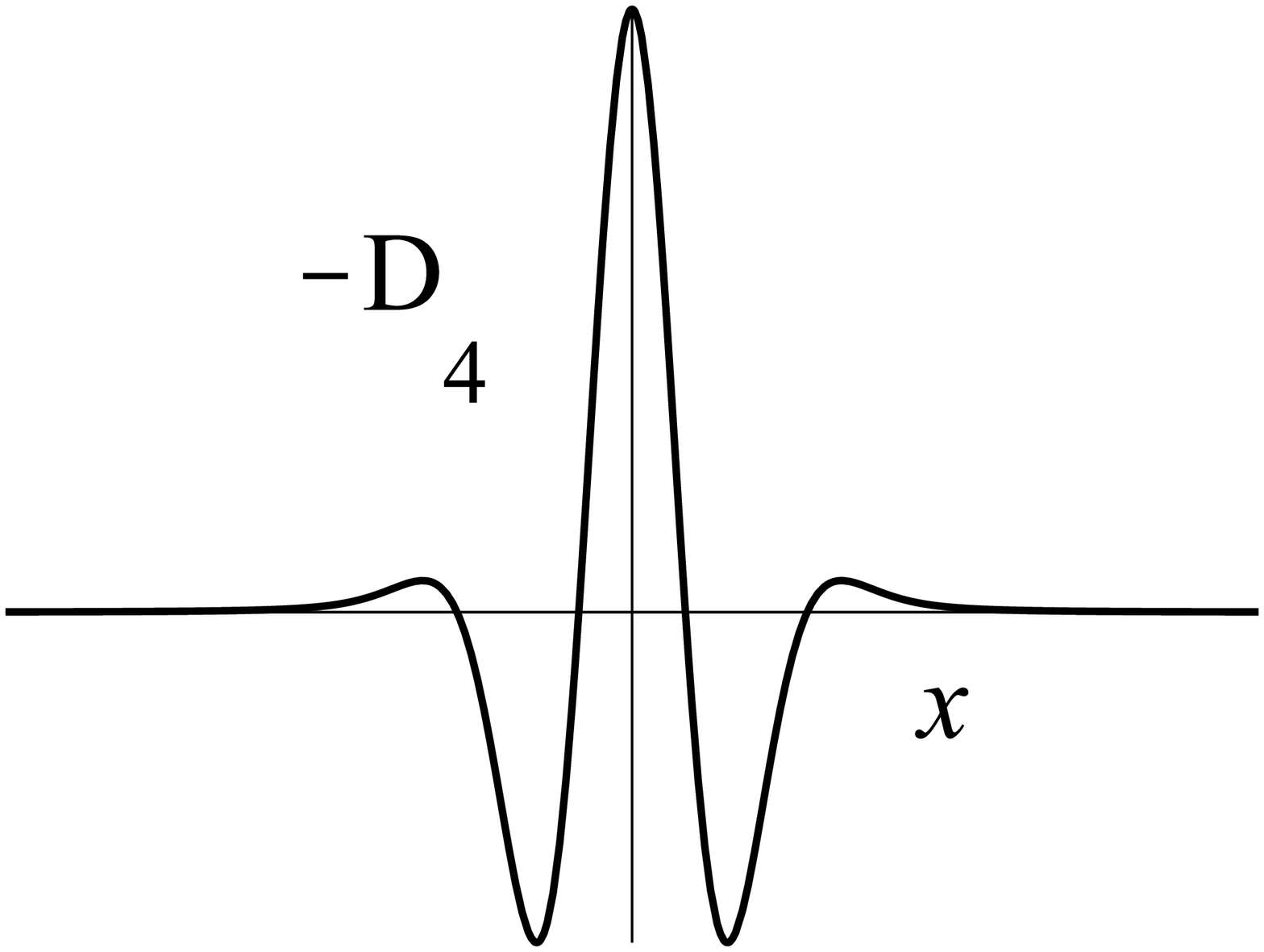}}
\caption{\label{susyt}
The SUSY type ladder relation between Dawson's eigenfunctions $D_n(x)$ and the wrong-sign Hermite eigenfunctions $\widetilde H_n(x)$.  The function $D_0(x)$ is missing, but can be found using the second order  differential equation (\ref{derkm1}) and the lowering operator (\ref{der1}).}
\end{center}\end{figure}

As one can see, equation (\ref{g-osc}) is the same as (\ref{derkm1}) when $k=1$ 
and $n=0$, i.e., 
the ground state equation of the wrong-sign Hermite eigenvalue problem, 
whose eigenfunction is just the Gaussian function
%
$
\widetilde H_0(x)= e^{-x^2}
$
%
and it is not Dawson's function!  
Now, $\widetilde H_0(x)$ and F$(x)$ are completely different, and their first 
derivatives are also different,
\begin{subequations}
\begin{align}
\label{derG}
 \widetilde H_0'(x) &= -2 x \widetilde H_0(x) \ ,\\ 
\label{derF}
 \mbox{F}'(x) &= -2 x \mbox{F}(x) +1 \, ,
\end{align} \label{dif1a}
\end{subequations}
however, they share the same second order differential equation.  The reason for 
this is that Dawson's function is the second solution of equation (\ref{g-osc}) 
when $n=0$, since starting with $\widetilde H_0(x)$, that solution is
\begin{equation}
f(x)= \widetilde H_0(x) \int_0^x 
\frac{e^{-\int^y 2z \, dz}}{\left(\widetilde H_0(y)\right)^2} \, dy
= F(x)  \ .
\end{equation}
It is even more interesting to see that, using 
eqs.(\ref{dder}), (\ref{g-osc}) and (\ref{tildehn}), we can find a SUSY like ladder relation between the wrong-sign Hermite eigenfunctions $\widetilde H_n(x)$, and the derivatives of Dawson's functions, which we shall hereafter call the Dawson's eigenfunctions,
$D_n(x)\equiv F^{(n-1)}(x)$, for $n=1,2,3,\ldots$, and hence $D_1(x)\equiv F(x)$.
This ladder relation is shown in Fig.\ref{susyt}.

The fact that $F(x)$ has one zero at $x=0$, while the Gaussian does not have any zero for all finite $x$, implies that in order that both sets of eigenfunctions cover the whole space of non-singular functions $f(x)$, there must exist an additional function $D_0(x)$, however enlarging the eigenvalue equations for Dawson's eigenfunctions.  This feature is the equivalent to the SUSY-QM procedure, where the zero-th order SUSY partner eigenfunction is missing.  To find the missing $D_0(x)$, we can see that eq.(\ref{dder}), in the case $k=0$, has two solutions, the constant solution, and the error function.  However, if we assume that the recursion relation (\ref{hnmenos}) gives rise to eq.(\ref{der1}), then, the (non-trivial) zero-th eigenfunction ought to be $D_0(x)=const$.

We say here that Dawson's eigenfunctions and the wrong-sign Hermite eigenfunctions possess a singular SUSY relation because ({\it i}) there is no QM problem associated, ({\it ii}) their relation did not arise from an operator procedure; moreover, they share the recurrence relations and the second order differential equation, ({\it iii}) there does not exist an associated SUSY parameter in this relation, and ({\it iv}) in order to cover the space of nonsingular functions of $x$, there must exist a zero-less eigenfunction $D_0(x)$, which in this case is the constant function.

\subsection{Dawson's function series approximation}

In order to find an analytical approximation for Dawson's function, let us first 
notice that the ratio $F(x)/x$ looks very much like a Gaussian but with 
very long tails.  This is the reason why Cody's rational approximations 
(\ref{codyf}) in terms of series of $x^2$ work pretty well.

\begin{figure}[H]\begin{center}
\subfigure{\includegraphics[width=0.485\textwidth]{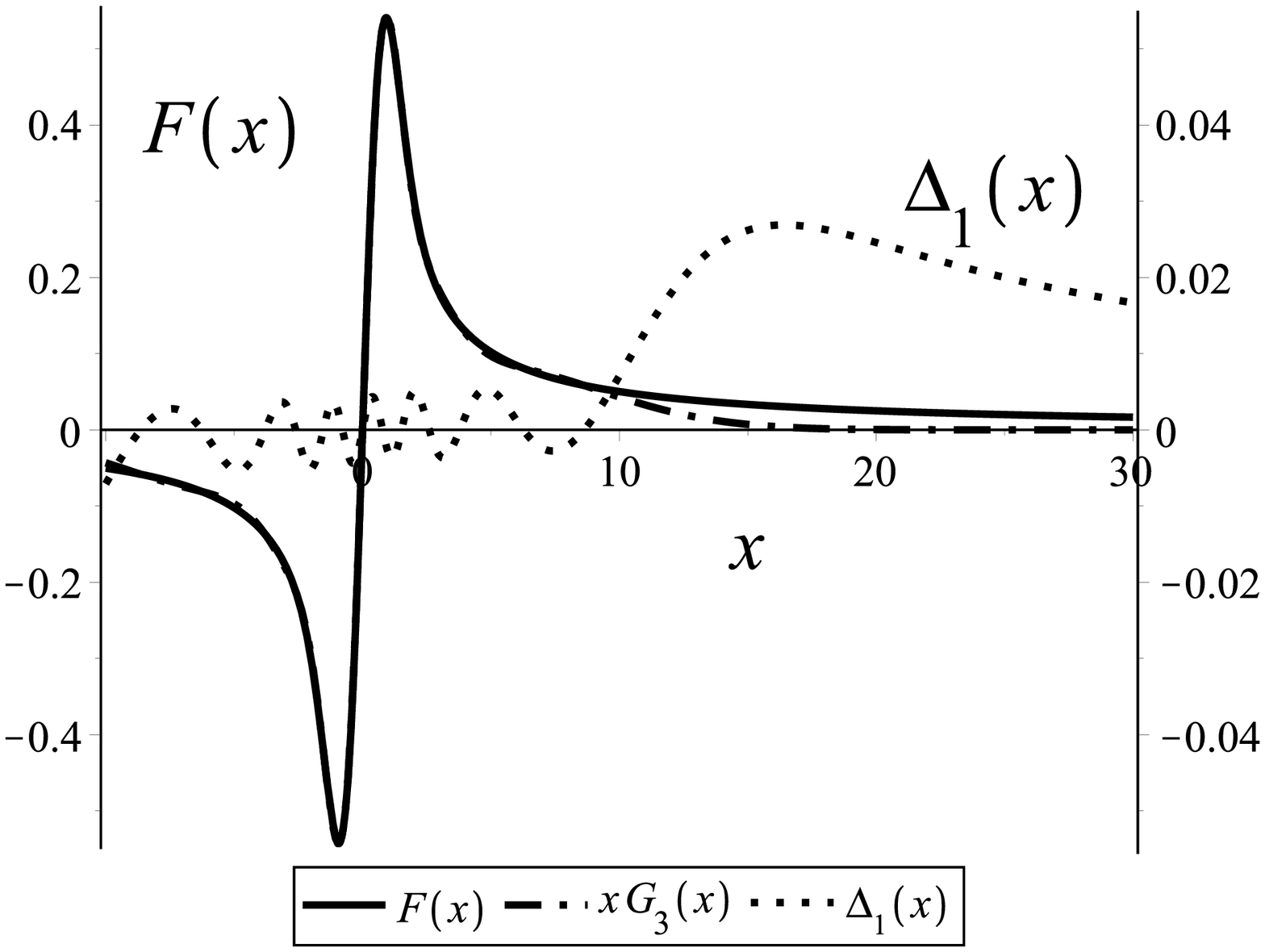}}
\subfigure{\includegraphics[width=0.485\textwidth]{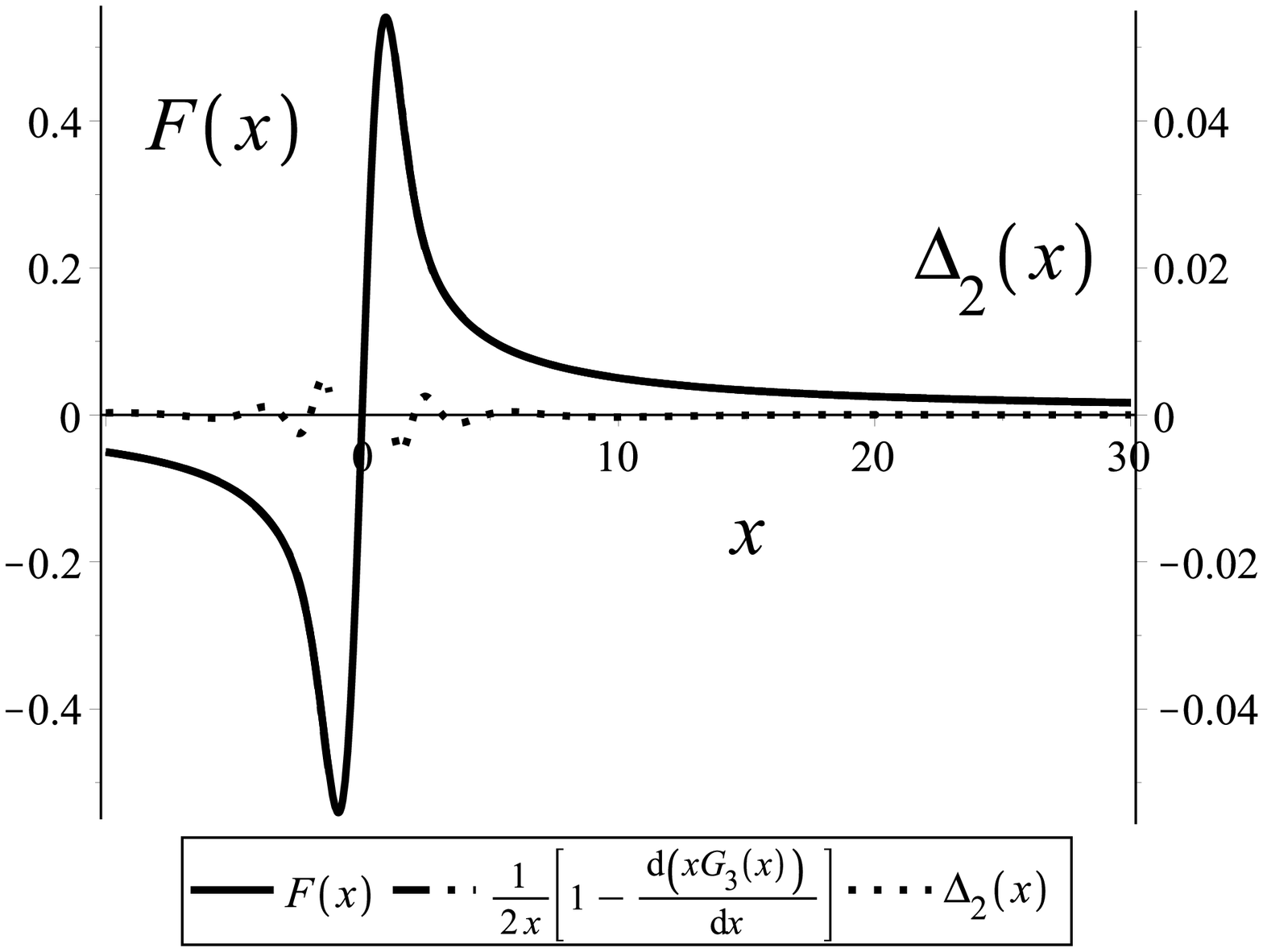} }
\caption{\label{dawff}
Left panel: The three Gaussians approximation to Dawson's function, where 
$G_3(x)=0.152 e^{-\frac{x^2}{2(1.804)^2}}
 + 0.805 e^{-\frac{x^2}{2(0.825)^2}} + 0.025 e^{-\frac{x^2}{2(5.536)^2}}$. 
Right panel: Fit to Dawson's function using the derivate of the three Gaussian approximation. The dotted curves, with values on the right $y$-axis, denote the difference $\Delta_1(x)$ and $\Delta_2(x)$, between Dawson's function and the approximations.
}
\end{center}\end{figure}

Given this similarity, a good approximation to Dawson's function can be realized 
in terms of a sum of Gaussians, all centered at the origin, and with different 
widths,
\begin{equation}\label{F-aprx}
F(x)= x  \sum_{i=1}^n a_i \, e^{-0.5 \, x^2/ \sigma_i^2}=x\, G_n(x).
\end{equation}
The fit of $F(x)$ with $n=3$ is shown in the left panel of Fig.\ref{dawff}.  
The left $y$-axis shows the function values, while the right $y$-axis denotes 
the difference $\Delta_1(x)=F(x)-xG_3(x)$.
It is interesting to note that for $|x|>2.5$, a better approximation is given by 
the relation
\begin{equation}\label{F-aprx2}
F(x)=\frac{1}{2x}\left( 1-\frac{d \left[xG_3(x)\right] }{dx} \right)
\end{equation}
which is based on eq.(\ref{der1}).  
The right panel of Fig.\ref{dawff} shows the difference $\Delta_2(x)$ between Dawson's function and this approximation, which is plotted there for $|x|\geq 1.28$, since it diverges when $|x|\to 0$.  Then, we can define a segmented analytical approximation for $F(x)$, the first one defined for $|x|\leq 2.397$, and the second one for $|x|> 2.397$, since they match at that point.

\section{Conclusion}

In this paper we have derived an analytical approximation for the error function, in terms of a finite series of elementary functions.   
This series is derived from the SUSY factorization of the Hamiltonian of a 
particle subject to a P\"oschl-Teller potential.  Considering another known SUSY factorization, we have defined the Dawson eigenfunctions as Dawson functions' derivatives, 
since they posses a singular SUSY type relation with the wrong-sign 
Hermite eigenfunctions.  It would be interesting to make a further analysis of this kind of SUSY like relations in other spectral problems, as this certainly broadens our knowledge of SUSY factorizations and the range of validity of the factorization parameters involved. Finally, we proposed a series expansion of Dawson's function in terms of Gaussian functions.

\section{Aknowledgements}

MAR wishes to thank CONACYT-Mexico for support for a sabbatical period at Fermilab.  RAO also wishes to thank CONACYT for a PhD scholarship.

\pagestyle{plain}

\end{document}